\documentclass[prb,aps,twocolumn,floatfix,longbibliography,showpacs]{revtex4-1}
\usepackage{amssymb}
\usepackage{amsmath}
\usepackage{graphicx}
\usepackage{bm}
\usepackage[colorlinks=true,linkcolor=blue,anchorcolor=red,citecolor=blue,urlcolor=blue]{hyperref}
\usepackage{bbm}
\usepackage{amsfonts}
\usepackage{booktabs}
\usepackage{mathrsfs}
\usepackage{epsfig}
\usepackage{graphicx}% Include figure files
\usepackage{dcolumn}% Align table columns on decimal point
\usepackage{slashed}
\usepackage{subfigure}
\usepackage{color, soul}
\usepackage{booktabs}

\begin{document}

\title{CH$_3$NH$_3$PbI$_3$/GeSe bilayer heterojunction solar cell with high performance}

\author{Guo-Jiao Hou$^{1}$}
%\affiliation{aaa, China}

\author{Dong-Lin Wang$^{2}$}
%\affiliation{bbb, China}

\author{Roshan Ali$^{1}$}

\author{Yu-Rong Zhou$^{3}$}

\author{Zhen-Gang Zhu$^{1,2}$}
\email{zgzhu@ucas.ac.cn}

\author{Gang Su$^{2,4}$}
\email{gsu@ucas.ac.cn}

\affiliation{$^{1}$ School of Electronic, Electrical and Communication Engineering, University of Chinese Academy of Sciences, Beijing 100049, China. \\
$^{2}$ Theoretical Condensed Matter Physics and Computational Materials Physics Laboratory, College of Physical Sciences, University of Chinese Academy of Sciences, Beijing 100049, China.\\
$^{3}$ College of Materials Science and Opto-Electronic Technology, University of Chinese Academy of Sciences, Beijing 100049, China.\\
$^{4}$ Kavli Institute of Theoretical Sciences, University of Chinese Academy of Sciences, Beijing 100190, China.}

%\date{\today }
\begin{abstract}
Perovskite (CH$_3$NH$_3$PbI$_3$) solar cells  have made significant advances recently. In this paper, we propose a bilayer heterojunction solar cell comprised of a perovskite layer combining with a IV-VI group semiconductor layer, which can give a conversion efficiency even higher than the conventional perovskite solar cell. Such a scheme uses a property that the semiconductor layer with a direct band gap can be better in absorption of long wavelength light and is complementary to the perovskite layer. We studied the semiconducting layers such as GeSe, SnSe, GeS, and SnS, respectively, and found that GeSe is the best, where the optical absorption efficiency in the perovskite/GeSe solar cell is dramatically increased. It turns out that the short circuit current density is enhanced $100\%$  and the power conversion efficiency is promoted $42.7\%$ (to a high value of $23.77\%$) larger than that in a solar cell with only single perovskite layer. The power conversion efficiency can be further promoted so long as the fill factor and open-circuit voltage are improved. This strategy opens a new way on developing the solar cells with high performance and practical applications.
\end{abstract}


\maketitle
% \tableofcontents

\section{Introduction}
% ---------------------------------------------------------------------------
Photovoltaic (PV) devices are used to convert the solar energy into electricity. How to choose appropriate materials is an indispensable issue  in design of PV devices. A number of properties are required for candidate PV materials, such as strong optical absorption over a wide range of light wavelength, good collection properties for carriers and low cost in large-scale applications, \cite{Reithmaier} and so on. Single crystal and polycrystalline Si have dominated the PV industry for long time owing to its mature manufacturing technology and abundant sources. Due to the indirect band gap of Si, the absorption of light has to be mediated by phonon-assisted processes, giving rise to limitations on the efficiency. Therefore, it is highly desired to look for better materials with direct band gap for a replacement of Si in solar cells, and much effort was paid over the years. To name a few, GaAs is a kind of semiconductor with direct band gap and shows good optical properties, and the efficiency in GaAs-based thin-film solar cell has been realized at $28.8\%$ \cite{Moon}.
However the high price is one of main obstacles to restrict massive applications of GaAs-based solar cells. The polycrystalline CdTe-based solar cell can absorb $90\%$ of the solar spectra with only $1$ $\mu$m thick \cite{Asim}, but the main obstacle  is the toxicity of cadmium, and thus, CdTe is not suitable for a large-scale and green PV application \cite{Shah}.
Organic solar cells have developed rapidly in recent decades due to technical developments of semiconducting polymers \cite{Nunzi,Hoppe}, which have many advantages, including low cost materials, high-throughput roll-to-roll production, mechanical flexibility and light weight \cite{Zhao}, but high stability and long cyclic lifetime are still big challenges. So, the mixed organic-inorganic halide perovskites (CH$_3$NH$_3$PbX$_3$, X=Cl,Br,I) jump into the stage and become booming PV materials in short time, where the conversion efficiency of perovskite-based solar cells rises rapidly from $3.8\%$ to $22.1\%$ in past few years \cite{Lin,Li}.

CH$_3$NH$_3$PbI$_3$ as an ambipolar semiconductor (n- or p-type) has intense light absorption and long diffusion length and lifetime of carriers. These advantages lead to a high power conversion efficiency (PCE) in CH$_3$NH$_3$PbI$_3$-based solar cells. Perovskite solar cells have various structures such as liquid-electrolyte dye-sensitized cells, mesoporous structure, planar n-i-p structure, planar p-i-n structure, HTL-free cells, and ETL-free cells \cite{Zuo}, etc. The planar p-i-n structure solar cell uses CH$_3$NH$_3$PbI$_3$ as an active layer to absorb light, while the layers above and beneath it play the role in conducting holes and electrons \cite{Lin}. The semi-transparent planar perovskite solar cells \cite{Fu} are also pursued for further improving the efficiency of perovskite solar cells.

The wavelength of light absorbed by perovskite solar cells ranges from $300$ nm to $800$ nm. This means the absorption energies are limited to the visible light, and most infrared light are actually wasted. Besides, the absorption coefficient of perovskites drops over $90\%$ at wavelengths beyond 650 nm \cite{650nm}. However, it is well known that the invisible ``near-infrared'' ($700-2500$ nm) radiation carries more than half of the power of sunlight \cite{Santamouris}.
Therefore, to improve the efficiency of energy utilization in a solar cell, recycling such wasted energies is an imperative choice. Unfortunately, it is impossible to reach this goal in a single-layer perovskite solar cell since the absorbers have fixed bandgaps that may bring limitations on absorbing energies.
%Therefore, it is an important way to extend light absorption range for the efficiency improvment of solar cells.

One idea is to combine more solar cells serially connected into a tandem (multi-junction (MJ)) solar cell. MJ solar cells are made of multiple p-n junctions from different semiconductors. Each p-n junction produces electric current in response to different range of wavelengths of solar spectrum, aiming to improve the total energy conversion efficiency.
Usually, there are two ways for MJ solar cell fabrication: series connection and parallel connection. For the former, a single current passes and only two electrodes are required. The single current is in fact determined by the weakest current generated in one sub-cell subject to the current match condition, constraining the performance of entire MJ solar cell. While for the latter, each sub-cell is connected by separate electrodes and the output from each sub-cell can be independently optimized. But it is not easy to fabricate the middle electrode in practice \cite{Bremner}.  
%
%\textcolor{blue}{Although the differences between the crystal structure or lattice constants of these two materials generally introduce defect states at the junction and these defect states would increase the recombination and limit the open circuit voltage, it is a effective method to improve absorption efficiency and easy to fabricate comparing to tandem solar cell. In addition, the improvement of preparation method and an appropriate intermediate layer can decrease the defect states effectively. }

We propose, in this work, a solar cell based on a bilayer p-n heterojunction serving as an active layer. One layer is selected to be perovskite, i.e. CH$_3$NH$_3$PbI$_3$, whose band gap is 1.5 eV, and the cut-off wavelength is $827$ nm according to $\lambda=\frac{hc}{E_{g}}$, where $h$ is the Planck constant, $c$ is the velocity of light, and $E_{g}$ is the band gap of the material.
%
%\textit{From the solar spectrum, in the infrared light region, with the wavelength increases, solar spectral radiance decreases. If the light of $800-1200nm$ can be effectively absorbed, we can say the solar cell utilized the most of infrared light. }
To recycle the unabsorbed energies of the light after penetrating into the perovskite layer, it is instructive to add another layer which should absorb the energies ranging in $800-1200$ nm.
%Then our aim to find such a material: it has better optical absorption properties in $800-1200nm$ and use it to compensate the lack of perovskite solar cell in light absorption.
To achieve this aim, we notice that the narrow-band-gap IV-VI group semiconductors constitute an important class of materials for photovoltaic applications \cite{Ii}. IV-VI group layered semiconducting compounds (e.g., SnS, SnSe, GeS, and GeSe) have attracted much attention due to their interesting optical and electrical properties \cite{El-Rahman,Deng}, which can be efficient PV materials with suitable bandgaps ranging from $0.5$ eV to $1.5$ eV \cite{Huang}. In particular, GeSe is a layered semiconductor that is predicted to bear a high chemical stability \cite{Ulaganathan} and can be used as low-cost components of photovoltaic cells \cite{Ii}. The experimental bandgap of GeSe is 1.1 eV \cite{Elkorashy}, implying that it is suitable for an active absorber in a solar cell. It was also shown to be promising materials for ultrathin-film flexible photovoltaic applications with high conversion efficiency, and can compete with organic and dye-sensitized solar cells \cite{Shi}. Therefore, we select these kind of IV-VI group semiconductors as another layer in our bilayer heterojunction solar cell structure. Here it is interesting to mention that we are pursuing to improve the efficiency of a \textit{single} p-n junction solar cell rather than a multi-junction solar cell. This may increase the output energy density, and there is no current match requirement for such a single p-n junction solar cell in our study.

\begin{figure}[tbp] %开始一张图
\begin{center} %开始图片居中
%\scalebox{0.18}{\epsfig{file=1_jiegou.eps}}%1.eps 为文件名 只需要将1 改成你的文件名即可 scalebox 里的数字用以调节图的大小
  \includegraphics[width=0.55\linewidth]{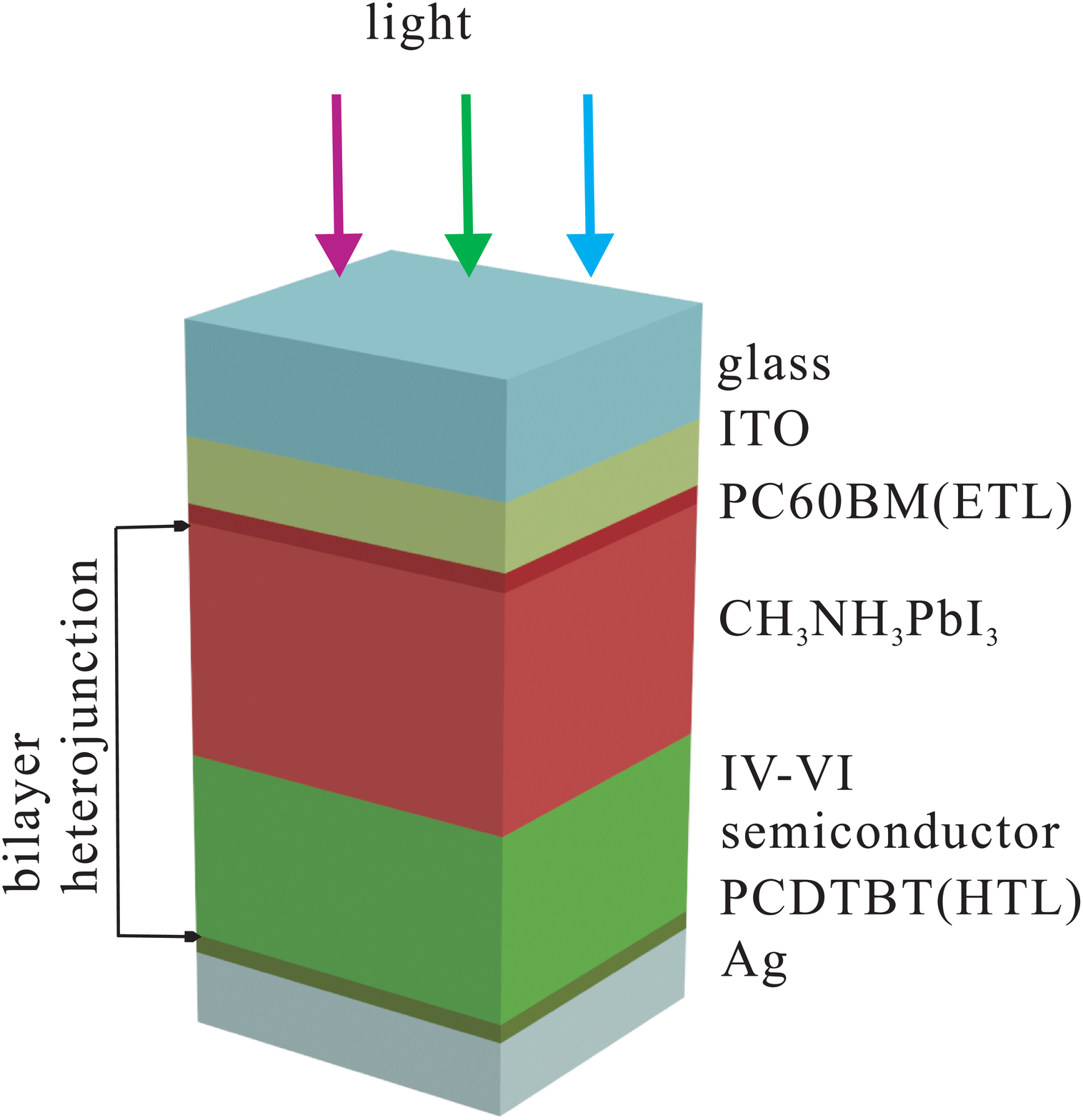}
\end{center} \caption{The schematic structure of the bilayer solar cell device. The active layer is a bilayer heterojunction comprised of CH$_{3}$NH$_{3}$PbI$_{3}$ layer and a IV-VI group semiconductor layer. A p-n junction is formed in the bilayer heterojunction. }
\label{structure}
\end{figure}%结束一张图

\section{optical absorption of the bilayer solar cell}

The structure of the bilayer solar cell we propose here is shown in Fig. \ref{structure}. The top layer is a glass with thickness 100 nm, and the second layer is 80 nm thick Indium tin oxide (ITO) which is taken as the anode of the solar cell. Perovskite (CH$_3$NH$_3$PbI$_3$) layer and IV-VI group semiconductor layer constitute a n-p heterojunction and are taken as active layers expected to absorb almost all light through the whole cell device. As CH$_3$NH$_3$PbI$_3$ as a semiconductor with wide band gap absorbs relatively short wavelength visible light, the layer of narrow-band-gap IV-VI group layered semiconductors under the CH$_3$NH$_3$PbI$_3$ layer can absorb the infrared light. The thickness of CH$_3$NH$_3$PbI$_3$ is taken to be 350 nm according to a previous experiment \cite{Lin}, and that of the IV-VI group semiconductor layer is set as 450 nm. On the top of CH$_3$NH$_3$PbI$_3$, there is an electronic transport layer (ETL): PC60BM (((6,6)-phenyl-C61-butyric acid methyl ester), and under the IV-VI group semiconductor layer, there is a hole transport layer (HTL): PCDTBT (poly(N-9$'$-heptadecanyl-2,7-carbazole-alt-
5,5-(4$'$,7$'$-di(thien-2-yl)-2$'$,1$'$, 3$'$-benzothiadiazole))) \cite{Lin}. The bottom layer is 100nm thick Ag as cathode.

\begin{figure} %开始一幅并列图
 \begin{center}
 % \scalebox{0.3}{\epsfig{file=2_xishou.eps}}
  \includegraphics[width=0.9\linewidth]{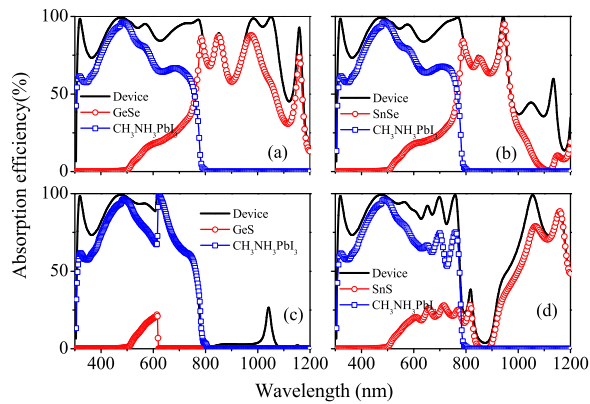}
  \end{center}
 \caption{The calculated absorption efficiency for the device of bilayer solar cell with (a) CH$_3$NH$_3$PbI$_3$/GeSe; (b) CH$_3$NH$_3$PbI$_3$/SnSe; (c) CH$_3$NH$_3$PbI$_3$/GeS; and (d) CH$_3$NH$_3$PbI$_3$/SnS.}
 \label{absorption}
 \end{figure}%结束一幅并列图

For IV-VI group semiconductor layers, here we consider GeSe, SnSe, GeS and SnSe due to their suitable direct bandgaps. The dielectric constants of these materials are taken from Ref. \onlinecite{Eymard}. According to the following equation
\begin{eqnarray}
n &=& \frac{1}{\sqrt{2}}\left(\epsilon_{1}+\left(\epsilon_{1}^{2}+\epsilon_{2}^{2}\right)^{\frac{1}{2}}\right)^{\frac{1}{2}}, \notag\\
k &=& \frac{1}{\sqrt{2}}\left(-\epsilon_{1}+\left(\epsilon_{1}^{2}+\epsilon_{2}^{2}\right)^{\frac{1}{2}}\right)^{\frac{1}{2}},
\label{nk}
\end{eqnarray}
we can get the refractive index of materials \cite{Fox}, where $\epsilon_{1}$ and $\epsilon_{2}$ are the real and the imaginary part of dielectric constants, respectively, and $n$ and $k$ are the real and imaginary part of the refractive index. Taking the refractive index as an input parameter,  we use the methods described in Refs. \onlinecite{wangDL2014,wangDL2015,Wang} to calculate the absorption efficiency. The other parameters are taken from Ref. \onlinecite{Wang}.

%%%%%%%%%%%%%%%%%%%%%%%%%%%%%%%%%%%%%%%%%%%%%%%%%%%%%%%%%%%%%%%%%%%%%%%%%%%%%%%%%%%%%%%%%%%%%%%%%%%%%%%%%%%%%%%%%%%%%%%%%%
 \begin{figure}[tb] %开始一张图
% \begin{center} %开始图片居中
% \scalebox{0.2}{\epsfig{file=3_nengji.eps}}%1.eps 为文件名 只需要将1 改成你的文件名即可 scalebox 里的数字用以调节图的大小
  \includegraphics[width=0.9\linewidth]{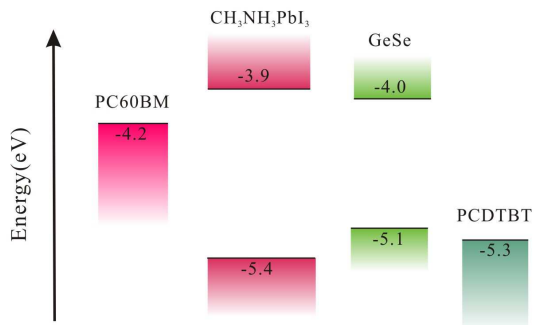}
%  \end{center}%结束图片居中
\caption{Schematic illustration of the energy levels of CH$_3$NH$_3$PbI$_3$, GeSe, the ``electron affinity" of PC60BM,  and the polymer interlayer ionization potential of PCDTBT. }
\label{bandprofile}
\end{figure}%结束一张图
%%%%%%%%%%%%%%%%%%%%%%%%%%%%%%%%%%%%%%%%%%%%%%%%%%%%%%%%%%%%%%%%%%%%%%%%%%%%%%%%%%%%%%%%%%%%%%%%%%%%%%%%%%%%%%%%%%%%%%%%%%

%%%%%%%%%%%%%%%%%%%%%%%%%%%%%%%%%%%%%%%%%%%%%%%%%%%%%%%%%%%%%%%%%%%%%%%%%%%%%%%%%%%%%%%%%%%%%%%%%%%%%%%%%%%%%%%%%%%%%%%%%%
\begin{table}[tb]
\centering  % 表居中
\caption{Bandgaps and corresponding cutoff wavelengths of GeSe, SnSe, GeS, and SnS. }
\begin{tabular}{lcccc}  % {lccc} 表示各列元素对齐方式，left-l,right-r,center-c
\hline\hline
Materials  &GeSe &SnSe &GeS &SnS \\ \hline   % \hline 在此行下面画一横线
Bandgap (eV) &1.1 \cite{Elkorashy}   &1.047 \cite{parenteau}  &1.61 \cite{Ii} &1.296 \cite{parenteau} \\         % \\ 表示重新开始一行
Cutoff wavelength (nm) & 1088 & 1184 & 770 & 957 \\        % & 表示列的分隔线
\hline
\end{tabular}

\label{bandgaps}
\end{table}
%%%%%%%%%%%%%%%%%%%%%%%%%%%%%%%%%%%%%%%%%%%%%%%%%%%%%%%%%%%%%%%%%%%%%%%%%%%%%%%%%%%%%%%%%%%%%%%%%%%%%%%%%%%%%%%%%%%%%%%%%%

We summarize the experimental bandgaps and corresponding cutoff wavelengths for bulk IV-VI group layered semiconductors, namely GeS, SnS, GeSe, and SnSe in Table \ref{bandgaps}. As the cutoff wavelengths in the range of $1000-1200$ nm are proper, GeSe and SnSe may be better choices for the present purpose.
This can be seen clearly in Fig. \ref{absorption} in which the absorption of the solar cell is calculated. ``Device" labels the entire absorption for the solar cell consisting of the bilayer heterojunction of CH$_3$NH$_3$PbI$_3$/IV-VI group semiconductor layers. The curves labeled by ``CH$_3$NH$_3$PbI$_3$" or IV-VI group semiconductors represent the contributions of these two materials in the cell device.
 Fig. \ref{absorption}(a) gives the absorption efficiency in the case of GeSe and CH$_3$NH$_3$PbI$_3$ as active layers. One may see that CH$_3$NH$_3$PbI$_3$ shows a strong absorption in the range of 300 nm to 800 nm in accordance with the bandgap. However, there is no absorption when the wavelength is larger than $800$ nm. On the contrary, GeSe layer starts to absorb light from a wavelength of $500$ nm and dramatically to a high absorption at around $800$ nm. When the wavelength is larger than $1200$ nm, the absorption decreases considerably. It is clear that the light absorption of the perovskite and GeSe layers are complementary to each other. Because of this property, the entire absorption of the bilayer heterojunction comprised of these two materials is perfect in the range of $300$ nm to $1200$ nm. In other words, the CH$_3$NH$_3$PbI$_3$/GeSe bilayer heterojunction solar cell can have a light absorption much better than the single CH$_3$NH$_3$PbI$_3$ layer solar cell.

Although the idea of heterojucntion solar cell is common and an improvement of absorption efficiency can be generally expected, the difficulty is to look for a proper material. One criterion to look for such a material is that the new material should be a direct semiconductor with proper bandgap. This rules out most semiconductors including Si, GaAs, etc. In this sense, we found GeSe can be an ideal candidate for such a purpose. It is noted that GeSe serving as an active layer in the application of single layer solar cells has been tested very recently.

Figs. \ref{absorption}(b)-(d) show the absorption efficiencies of the bilayer heterojunction solar cells with perovskite and SnSe, GeS and SnS, respectively.  It is seen that SnSe has also a good absorption in the range of 800 nm to 1200 nm. In comparison to GeSe, although they belong to the same category of materials, GeS and SnS do not show much improvements in absorption as compared with the two selenides. We therefore conclude that GeSe may be the best materials for our bilayer heterojunction solar cell. In the following we shall focus only on GeSe.

\begin{figure}[tb] %开始一张图
 \begin{center} %开始图片居中
% \scalebox{0.15}{\epsfig{file=4_nengdai.eps}}%1.eps 为文件名 只需要将1改成你的文件名即可 scalebox里的数字用以调节图的大小
   \includegraphics[width=1\linewidth]{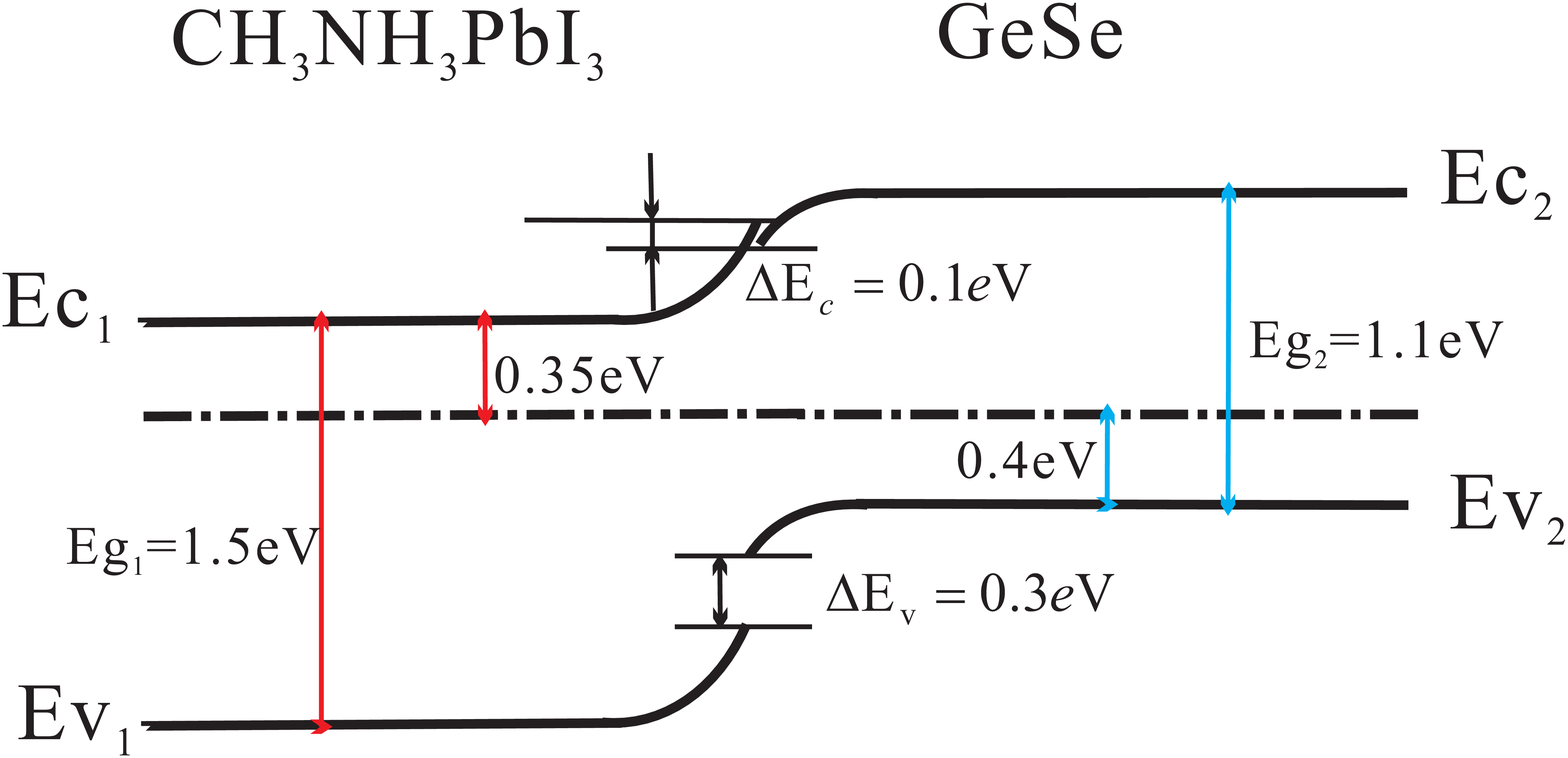}
 \end{center}%结束图片居中
\caption{The band profile of CH$_3$NH$_3$PbI$_3$/GeSe heterojunction.}
\label{equilibriumband}
\end{figure}%结束一张图

\section{electrical properties and efficiency of the bilayer solar cell}

The solar cell must satisfy electron transfer condition, so the work functions of every layer need to be suitable. For this purpose, in Fig. \ref{bandprofile}, we show the ionization potentials (IPs) of  PC60BM and  PCDTBT, as well as conduction band and valence band of CH$_3$NH$_3$PbI$_3$ and GeSe. PC60BM and  PCDTBT are commonly taken as ETL and HTL, respectively \cite{Lin,Ziaur}.
ETL  and HTL are thin n- or p-type inserted layers between cathode or anode and active layers, which can assist carrier extraction efficiently by reducing the eletrode-junction potential barrier.
The IP of PC60BM (-4.2eV) is lower than the conduction band edge of CH$_3$NH$_3$PbI$_3$ (-3.9 eV). Therefore it is easy for electrons to transfer from CH$_3$NH$_3$PbI$_3$ to PC60BM, but it is a barrier for holes. On the other hand, although the valence band edge of GeSe (-5.1eV) is slightly higher than the IP of PCDTBT (-5.3eV), such a small mismatch has little effect on hole transportation.

\begin{figure}[tb] %开始一张图
 \begin{center} %开始图片居中
% \scalebox{0.4}{\epsfig{file=5_G+U.eps}}%1.eps 为文件名 只需要将1改成你的文件名即可 scalebox里的数字用以调节图的大小
   \includegraphics[width=0.9\linewidth]{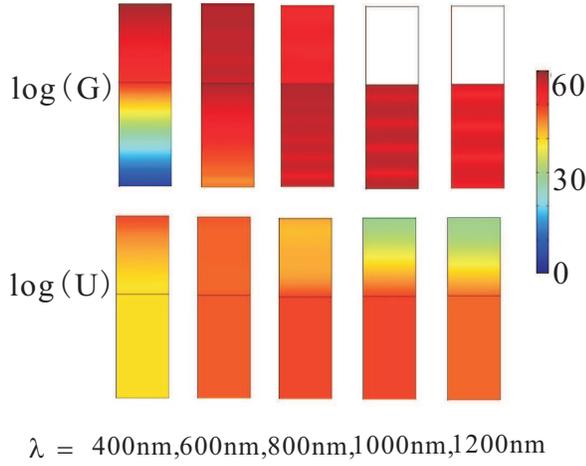}
 \end{center}%结束图片居中
\caption{ Profiles of generation rate (G)  and recombination rate (U) of carriers in the bilayer solar cell with $450$ nm GeSe layer and $350$ nm perovskite layer. The top layer (above the black line) is the perovskite layer, while the bottom layer (below the black line) represents GeSe layer.}
\label{GU}
\end{figure}%结束一张

The bilayer heterojunction of perovskite (CH$_3$NH$_3$PbI$_3$) and GeSe can lead to dissociation of electron-hole pairs in the solar cell. The equilibrium band profile of heterojunction is shown in Fig. \ref{equilibriumband}.  The position of Fermi level is determined by the relationship between doping concentration and the Fermi energy.  When the heterojunction occurs between these two semiconductors, there would be a conduction band step $\Delta E_c$ and valence band step $\Delta E_c$. The value of $\Delta E_c$ and $\Delta E_v$ are given by \cite{Milnes}

\begin{eqnarray} %开始一个公式
\Delta E_c &=& \chi_{1}- \chi_{2}, \notag\\
%\end{eqnarray}
%\begin{eqnarray}
\Delta E_v &=& (E_{g1}-E_{g2})-(\chi_{1}- \chi_{2}),
\label{DEcEv}
\end{eqnarray} %结束一个公式
where $\chi_{1}$ ($\chi_{2}$) is the electron affinity of CH$_3$NH$_3$PbI$_3$ (GeSe), and $E_{g1}$ ($E_{g2}$) is the bandgap of  CH$_3$NH$_3$PbI$_3$ (GeSe). The p-type GeSe and n-type CH$_3$NH$_3$PbI$_3$ have the same Fermi level and a small barrier 0.1 eV due to the offset of the conduction band edges. This barrier acts like a Shocttly barrier. However, the effect on photocurrent collection is negligible for a barrier lower than 0.4 eV \cite{Nelson}. Likewise, there is a small step in the valence band (0.3 eV) which may facilitate the hole transport from CH$_3$NH$_3$PbI$_3$ to GeSe.

%\textcolor{blue}{An important problem should be considered is the recombination at the interface of the perovskite and GeSe. At the GeSe surface,there are some dangling bonds because the valencies of surfaces atoms are unsatisfied. When the perovskite is deposited on top of GeSe surface, perovskite molecule does not chemically interact with the GeSe dangling bonds and some dangling bonds remain on the heterojunction interface. Dangling bonds may cause high density of defect states and increase the recombination of carriers. One way to reduce the interface defect density is to passivate the GeSe dangling bonds by an intermediate layer, deposited between GeSe and perovskite. The passivating layer is deposited first to consume the dangling bonds at the GeSe surface, and then the perovskite layer is added on top of the passivating layer to engineer the desired electrical properties of the heterojunction.}

Although the optical properties already give some hints, it is still necessary to study the electrical properties for a better conversion efficiency. The electrical properties of a solar cell can be reflected by the short circuit current density, open circuit voltage, fill factor and conversion efficiency. To obtain these quantities, we should first calculate the generation and recombination rates of carriers in the solar cell.

%%%%%%%%%%%%%%%%%%%%%%%%%%%%%%%%%%%%%%%%%%%%%%%%%%%%%%%%%%%%%%%%%%%%%%%%%%%%%%%%%%%%%%%%%%%%%%%%%%%%%%%%%%%%%%%%%%%%%%%%%%%%%%%%%%%%%%%%%
\begin{figure}[tb]
\includegraphics[width=0.9\linewidth]{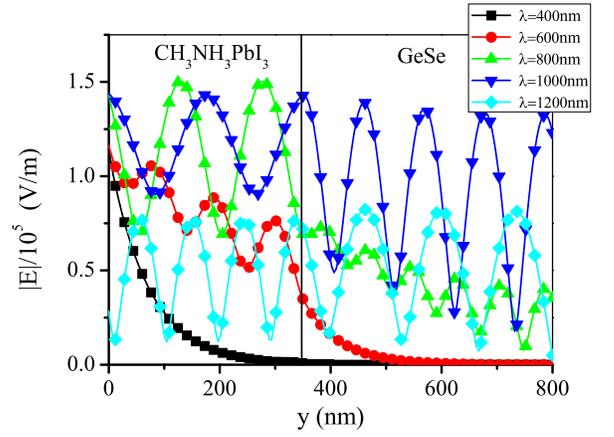}
%\vspace{-0.5cm}
\caption[]{The spatial distribution of electric field ($E$) in the perovskite/GeSe bilayer solar cell, where $y$ is the spatial distance from the top to bottom layer. }\label{Efield}
\end{figure}
%%%%%%%%%%%%%%%%%%%%%%%%%%%%%%%%%%%%%%%%%%%%%%%%%%%%%%%%%%%%%%%%%%%%%%%%%%%%%%%%%%%%%%%%%%%%%%%%%%%%%%%%%%%%%%%%%%%%%%%%%%%%%%%%%%%%%%%%%

We suppose all the absorbed photons can create excess carriers in the solar cell. The generation rate (G) can be given by an optical simulation. Recombination of carriers is the loss of an electron or hole through the decay to a lower energy state \cite{Nelson}. Recombination rate (U) can be obtained by an electrical simulation. The details of G and U calculations can be found in Sec. \ref{methods section}. Fig. \ref{GU} shows the spatial profiles of G and U when the incident light wavelength is $400$, $600$, $800$, $1000$ and $1200$ nm, respectively.
The top indicates CH$_3$NH$_3$PbI$_3$ and the bottom represents GeSe, as in Fig. \ref{GU}. In the top layer (perovskite), G is large for short wavelength (say $\lambda=400$ and $600$ nm) of incident light, corresponding to a wider bandgap, and it gradually decreases as the wavelength of incident light increases ($\lambda>600$ nm). When $\lambda$ is larger than $1000$ nm, the energy of incident light cannot excite excess carriers so that G in the perovskite layer tends to vanish. For the GeSe layer, things go different. For $\lambda=400$ nm, as the energy of incident light is larger than the bandgap of GeSe, G is not zero. However, it may excite the transition to higher bands, and reduce the generation magnitude. For longer $\lambda$, G in GeSe layer increases. When $\lambda>800$ nm, GeSe plays a major role in the device absorption. It is consistent with the optical absorption shown in Fig \ref{absorption}(a). The generation profile shows clearly that the two materials play complementary roles in light absorption of the bilayer cell structure, probably leading to a higher efficiency.
The recombination rate U behaves in a way similar to the generation rate G, as the recombination rate is proportional to the excess carriers generated by the incident light.

In addition, we observe that G has weak periodic fluctuations in each layer, where the periods of fluctuation depend on $\lambda$. To understand this periodicity, we look at the formula $G(\lambda)\propto \varepsilon''|E(\lambda)|^2$ (see Eq. (\ref{G}) in Sec. \ref{methods section}), where $\varepsilon''$ is the imaginary part of permittivity of materials, and $|E(\lambda)|$ is the magnitude of electric field. The oscillation of the generation rate G is due to the oscillation of electric field. We show the spatial dependence of $|E(\lambda)|$ in Fig. \ref{Efield}. For $\lambda=400$ nm, the electric field decays exponentially from perovskite layer to GeSe layer. For $\lambda=600$ nm, the electric field oscillates in perovskite layer, while it decays in GeSe layer. For longer wavelengths ($\lambda>1000$ nm), the electric field oscillates all over the bilayer. This oscillation of the magnitude of electric field reflects perfectly the periodicity of the generation rate G.
With the increase of wavelength, the penetration depth of light increases so that more light can arrive at a deeper position, even reaching the bottom Ag layer where the light can be reflected then. The vanishing generation rate G in perovskite layer for $\lambda>1000$ nm is not a consequence of $|E|$, but a vanishing $\varepsilon''$ for a long wavelength (with corresponding energies less than the bandgap of perovskite).

%%%%%%%%%%%%%%%%%%%%%%%%%%%%%%%%%%%%%%%%%%%%%%%%%%%%%%%%%%%%%%%%%%%%%%%%%%%%%%%%%%%%%%%%%%%%%%%%%%%%%%%%%%%%%%%%%%%%%%%%%%%%%%%%%%%%%%%%%%%%%%%%
\begin{figure}[tb]
\begin{center}
  \includegraphics[width=0.45\textwidth]{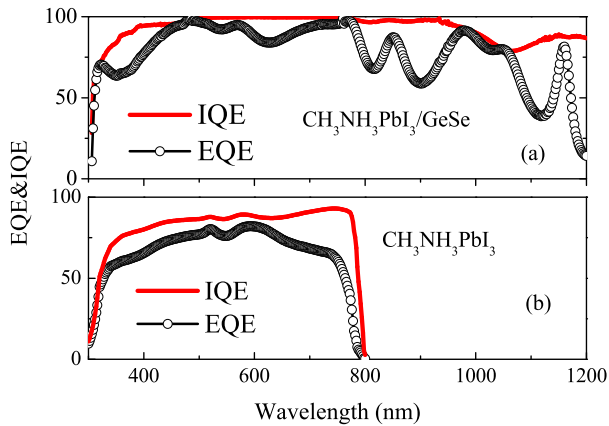}
\end{center}
\caption{(a) EQE and IQE of perovskite/GeSe bilayer solar cell; (b) EQE and IQE of perovskite single layer solar cell for a comparison.}
\label{EQEIQE}
\end{figure}%
%%%%%%%%%%%%%%%%%%%%%%%%%%%%%%%%%%%%%%%%%%%%%%%%%%%%%%%%%%%%%%%%%%%%%%%%%%%%%%%%%%%%%%%%%%%%%%%%%%%%%%%%%%%%%%%%%%%%%%%%%%%%%%%%%%%%%%%%%%%%%%%%

We solve the Poisson's equation and carrier transport equations self-consistently \cite{X Li}, and obtain the internal quantum efficienc (IQE) and external quantum efficiency (EQE), as shown in Fig. \ref{EQEIQE}(a). For a comparison, we also calculate EQE and IQE of the single layer perovskite solar cell [Fig. \ref{EQEIQE}(b)]. It can be seen that the EQE and IQE of perovskite/GeSe bilayer solar cell distribute in the range of $300$ nm to $1200$ nm of the solar spectra, while those of the single layer perovskite cell only distribute in the range of $300-800$ nm.

%%%%%%%%%%%%%%%%%%%%%%%%%%%%%%%%%%%%%%%%%%%%%%%%%%%%%%%%%%%%%%%%%%%%%%%%%%%%%%%%%%%%%%%%%%%%%%%%%%%%%%%%%%%%%%%%%%%%%%%%%%%%%%%%%%%%%%%%%%%%%%%%
 \begin{figure}[tb] %开始一张图
\begin{center}
 %\scalebox{0.3}{\epsfig{file=7_jv.eps}}%1.eps 为文件名 只需要将1改成你的文件名即可 scalebox里的数字用以调节图的大小
    \includegraphics[width=0.9\linewidth]{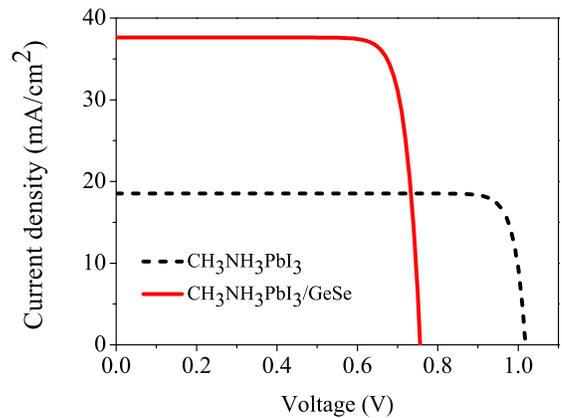}
\end{center} \caption{J-V curves of perovskite/GeSe bilayer and perovskite single layer solar cell for a comparison. }
\label{JV}
\end{figure}%结束一张图
%%%%%%%%%%%%%%%%%%%%%%%%%%%%%%%%%%%%%%%%%%%%%%%%%%%%%%%%%%%%%%%%%%%%%%%%%%%%%%%%%%%%%%%%%%%%%%%%%%%%%%%%%%%%%%%%%%%%%%%%%%%%%%%%%%%%%%%%%%%%%%%%

The resistance of solar cells is important in evaluating the efficiency, which, however, is difficult to calculate via a microscopic theory due to complicated scattering mechanisms. Fortunately, an equivalent circuit for a solar cell always contains the parasitic series ($R_{s}$) and shunt ($R_{sh}$) resistance. The shunt resistance is caused by the leakage current of the device generally due to the non-ideal devices \cite{Nelson}. The series resistivity has two main sources. One is from the contact resistances due to multi-interfaces between heterojunction, the active layer and the contacts, etc. The other is from the drift current density \cite{neamen}, i.e.  $J_{drf}=e(\mu_nn+\mu_pp)E=\sigma E$, where $\mu_{n(p)}$ is the mobility of electrons (holes), $n(p)$ is the electron (hole) concentration, and $\sigma$ is the conductivity of the semiconductor material. It leads us to the drift resistivity
\begin{eqnarray} %开始一个公式
\rho=\frac{1}{\sigma}=\frac{1}{q(\mu_nn+\mu_pp)}.
\label{resistivity}
\end{eqnarray} %结束一个公式
In a doped material, the drift resistivity is dominated by the majority carrier concentration and mobility. The drift resistance is therefore determined by the drift resistivity multiplying the material thickness. In terms of the parameters listed in Table \ref{parameters}, we obtain the drift resistance of GeSe (CH$_3$NH$_3$PbI$_3$) as $0.16$ $\Omega\cdot$cm$^2$ ($2.22$ $\Omega\cdot$cm$^2$). For the contact resistance, it is hard to know the value precisely. The total series resistance should be a sum of the total drift resistance and the contact resistance in the device. Since the contact resistivity is hard to obtain precisely, we assume that the series resistances of GeSe layer and perovskite layer are the same and equal to that of a single perovskite layer studied in Ref. \onlinecite{Wang} in which the perovskite solar cell is explored and its resistance is $6.4$ $\Omega\cdot$cm$^2$. In this way, the total series resistance of the bilayer solar cell is $12.8$ $\Omega\cdot$cm$^2$. The shunt resistance of the bilayer cell is assumed to be the same as that of a perovskite solar cell in Ref. \onlinecite{Wang}, which is 1.6 $k\Omega\cdot$cm$^2$.

Through the electrical simulation, $J-V$ curve for the bilayer solar cell is presented  in Fig. \ref{JV}. For a comparison, we include the $J-V$ curve  for typical perovskite solar cell as well. We also obtain the short circuit current density ($J_{sc}$), the open circuit voltage ($V_{oc}$), the fill factor ($FF$) and the conversion efficiency, as shown in Table \ref{electricalproperties}. Compared with the conventional perovskite solar cell, the $J_{sc}$ is improved significantly in the present bilayer solar cell, which is supported by the values of $37.62$ mA/cm$^2$ for the bilayer solar cell and 18.53 mA/cm$^2$ for the single-layer solar cell. This rise of $J_{sc}$ stems from more absorption of long wavelength light by introducing GeSe layer in the solar cell. The $V_{oc}$ of the bilayer solar cell is $0.76$ V, which is smaller than $1.02$ V of the typical perovskite solar cell. The $FF$ of bilayer solar cell is slightly smaller than that of the perovskite  solar cell since GeSe layer leads to a larger series resistance. By combining these parameters together, we obtain the efficiency of the perovskite/GeSe bilayer solar cell to be 23.77\%, which is much higher than 16.66 \% of the  conventional perovskite solar cell. The main reason for such a dramatic improvement is from a big increase of the short circuit current density. This observation demonstrates that our strategy of improving the output energy density in a single solar cell is reasonable. This may pave a way to improve the conversion efficiency further.

Here we would like to remark that, recently, the efficiency of conventional perovskite solar cell has been much improved in experiments, which can be over e.g. $20\%$ \cite{singlePerovskitesolarcell}. In practice, a 350 nm thick perovskite layer might not be sufficient for planar solar cells \cite{ZLiu}. However, in our study, we aim to demonstrate the applicability of the perovskite/GeSe bilayer solar cell, and fix the thickness of perovskite layer to be 350 nm, leading to the calculated efficiency ($16.66\%$) of the perovskite solar cell is lower than those reported very recently \cite{singlePerovskitesolarcell}. If the thickness of the perovskite is increased, the efficiency should be further enhanced.

We have some remarks here. Firstly, the hetero-junction structure may bring us more complexities. Extra defects may be introduced around the interface, which could increase the recombination rate and reduce the open circuit voltage. Therefore, they should be minimized in realistic applications. This might be achieved at some extent by using advanced nanotechnologies that may lead us atomic-resolved ideal interfaces. For the possible defects of dangling bonds, an inter-layer may be added to passivate them. Secondly, the crystal mismatch should be also considered. In our case, the crystal constants of the c-axis of perovskite and of the a-axis of GeSe are $a_{c,perovskite}=12.66${\AA}\cite{Mosconi} and $a_{a,GeSe}=10.862${\AA}\cite{Makinistian} respectively. The mismatch of the crystal constants is 14.2\% from $(a_{c,perovskite}-a_{a,GeSe})/a_{c,perovskite}$ and not so large. Moreover, in realistic applications, inserting buffer layers may be an alternative way to soften the mismatch. Thirdly, the carrier separation in the internal of each layer should be taken into account as well. For single p-n junction solar cells, the illumination of light can generate electron-hole pairs, which can be separated and swept out to produce the photocurrent due to the built-in electric field in the depletion region \cite{neamen}. Therefore, the thickness of the depletion region is quite important for a single p-n junction solar cell. In the present case, the calculated depletion lengths in GeSe and perovskite are about $8$ $\mu$m  and $300$ nm, and the thicknesses of the GeSe layer and the perovskite layer are $450$ nm and $350$ nm, respectively. We may state that the separation mechanism in the depletion region can work well in our device.

%%%%%%%%%%%%%%%%%%%%%%%%%%%%%%%%%%%%%%%%%%%%%%%%%%%%%%%%%%%%%%%%%%%%%%%%%%%%%%%%%%%%%%%%%%%%%%%%%%%%%%%%%%%%%%%%%%%%%%%%%%%%%%%%%%%%%%%%%%%%%%%%%%%%%%%%%%
\begin{table}[tb]
\centering  % 表居中
\caption{The relevant electrical quantities of perovskite/GeSe bilayer solar cell and conventional perovskite solar cell.}
\begin{tabular}{lccccc}  % {lccc} 表示各列元素对齐方式，left-l,right-r,center-c
\hline\hline
      Device type           & $J_{sc}$(mA/cm$^2$)    &$V_{oc}$(V)      &$FF$ ($\%$)  &Efficiency($\%$) \\ \hline   % \hline 在此行下面画一横线
 perovskite/GeSe            &37.62                   &0.76                   &83.14             &23.77 \\         % \\ 表示重新开始一行
 perovskite                &18.53                   &1.02                   &88.15              &16.66 \\
\hline
\end{tabular}

\label{electricalproperties}
\end{table}
%%%%%%%%%%%%%%%%%%%%%%%%%%%%%%%%%%%%%%%%%%%%%%%%%%%%%%%%%%%%%%%%%%%%%%%%%%%%%%%%%%%%%%%%%%%%%%%%%%%%%%%%%%%%%%%%%%%%%%%%%%%%%%%%%%%%%%%%%%%%%%%%%%%%%%%%%%

%%%%%%%%%%%%%%%%%%%%%%%%%%%%%%%%%%%%%%%%%%%%%%%%%%%%%%%%%%%%%%%%%%%%%%%%%%%%%%%%%%%%%%%%%%%%%%%%%%%%%%%%%%%%%%%%%%%%%%%%%%%%%%%%%%%%%%%%%%%%%%%%%%%%%%%%%%
\begin{table}[tb]
\caption{The relevant parameters of GeSe and CH$_3$NH$_3$PbI$_3$ materials, where $m_0$ is the bare mass of electron.}
\begin{tabular}{lcl}
\hline\hline
       Properties                        & GeSe                   & CH$_3$NH$_3$PbI$_3$\\
\hline
doping (cm$^{-3}$)                      & 3.54$\times10^{12}$        & 9.83$\times10^{13}$\\
bandgap (eV)                    & 1.14                      & 1.5  \\
electron mobility (m$^2/v\cdot s$)              & 0.05                     & 1$\times10^{-4}$\\
hole mobility (m$^2/v\cdot s$)                 &0.05            &$1\times10^{-4}$\\
electron effective mass ($kg$)         & 0.3$m_0$                  & 2$m_0$\\
hole effective mass ($kg$)            & 0.78$m_0$                  & 0.6$m_0$\\
electron lifetime ($s$)               & 0.01                       & 9.6$\times10^{-9}$ \\
 hole lifetime ($s$)                 & 0.01                    & 9.6$\times10^{-9}$ \\
relative permittivity          & 6.3                      &25 \\
\hline
\end{tabular}

\label{parameters}
\end{table}
%%%%%%%%%%%%%%%%%%%%%%%%%%%%%%%%%%%%%%%%%%%%%%%%%%%%%%%%%%%%%%%%%%%%%%%%%%%%%%%%%%%%%%%%%%%%%%%%%%%%%%%%%%%%%%%%%%%%%%%%%%%%%%%%%%%%%%%%%%%%%%%%%%%%%%%%%%

\section{Methods \label{methods section}}

Optical properties can be obtained by solving Maxwell equations in a finite element method (FEM) software package \cite{Li}. We consider both the transverse electric (TE) and transverse magnetic (TM) polarized incident light and take the average of these two modes. The AM1.5 global spectrum is used as incident solar spectrum in optical simulation. Refractive index and extinction coefficient of materials are input parameters of all layers in a solar cell and then the optical absorption efficiency of the device can be obtained by
\begin{eqnarray} %开始一个公式
A\left(\lambda\right)=\int\frac{\omega\varepsilon^{''}|E\left(\lambda\right)^{2}|}{2}dV,
\end{eqnarray} %结束一个公式
where $E\left(\lambda\right)$ is the distribution of the electric field intensity in each layer of solar cell as a function of wavelength, $\omega$ is the angular frequency of incident light, $\varepsilon^{''}$ is the imaginary part of permittivity of materials, $dV$ is the integral unit for volume of the solar cell. The density of photo-generated current ($J_{\text{G}}$) is given by \cite{Yu}
\begin{eqnarray} %开始一个公式
J_{G}=q\int\frac{A(\lambda)P_{am1.5}(\lambda)\lambda}{hc}d\lambda,
\end{eqnarray} %结束一个公式
where $q$ is the charge of an electron, $c$ is the speed of light, $h$ is the Planck constant, $P_{am1.5}(\lambda)$ is the spectral photon flux density in solar spectrum (AM 1.5). Light absorption leads to the generation of excess carriers with a generation rate expressed by
\begin{eqnarray}
G\left(\lambda\right)=P_{am1.5m}\left(\lambda\right)\frac{\varepsilon^{''}|E\left(\lambda\right)|^2}{2\hbar},
\label{G}
\end{eqnarray}
were $\hbar$ is the reduced Planck constant.
The electrons in excited states can relax to unoccupied states with lower energies at a recombination rate given by
\begin{eqnarray} %开始一个公式
U=\frac{np-n_{i}^{2}}{\tau_n(n+n_t)+\tau_p(p+p_t)},
\label{U}
\end{eqnarray} %结束一个公式
%
%$G\left(\lambda\right)$ can be as the parameter which combines optical simulation with electrical simulation in solar cell calculation.
where $n_i$ is the intrinsic carrier concentration of semiconductor, $\tau_n$ and $\tau_p$ are lifetimes of electrons and holes, respectively, $n_t$ ($p_t$) is electron (hole) concentration in a trapped energy level. The electrical properties of solar cell is accomplished by solving Poisson's equation and carrier transport equations \cite{Ferry,Deceglie}. The electrostatic potential ($\Phi$) is determined by Poisson's equation, and electron and hole concentrations can be obtained by
\begin{eqnarray}
\nabla^{2}\Phi &=& \frac{q}{\epsilon_{0}\epsilon_{r}}(n-p-N), \notag\\
\nabla(-D_{n}\nabla n+n\mu_{n}\nabla\Phi) &=& G\left(\lambda\right)-U\left(\lambda\right), \notag\\
\nabla(-D_{p}\nabla p-p\mu_{p}\nabla\Phi) &=& G\left(\lambda\right)-U\left(\lambda\right),
\label{transportequations}
\end{eqnarray}
where $N$ is the doping concentration, $\epsilon_{0}$ and $\epsilon_{r}$  are the vacuum and relative permittivity, respectively,  $D_{n}(D_{p})$ is the electron (hole) diffusion coefficient. Moreover, it is important to choose suitable initial values of parameters in the simulation as input. The initial potential is set by  \cite{X Li}
\begin{eqnarray} %开始一个公式
\Phi_{init}=\frac{k_{B}T}{q}\text{arcsinh}\left(\frac{N}{2n_{i}}\right),
\end{eqnarray} %结束一个公式
where $k_{B}$ is the Boltzmann constant, and $T$ is temperature. The initial carrier concentrations are set by
\begin{eqnarray} %开始一个公式
p_{init} &=& \sqrt{n_{i}^{2}+N^{2}/4}-N/2, \notag\\
n_{init} &=& \sqrt{n_{i}^{2}+N^{2}/4}+N/2.
\end{eqnarray} %结束一个公式
The parameters used in this device are presented in Table \ref{parameters}, which are taken from Refs. \onlinecite{Vodenicharov,Elkorashy,Wang,Xue}.

\section{Conclusion \label{Sec:conclusion}}
To improve the power conversion efficiency of the typical perovskite solar cell, we studied a bilayer solar cell with perovskite CH$_3$NH$_3$PbI$_3$ and IV-VI group semiconductor materials forming a heterojunction. Four semiconductors are considered in this paper, say, GeSe, SnSe, GeS, and SnS. The bandgaps of these four materials are suitable for the absorption of long wavelength light, which might have a complementary contribution to the perovskite that is almost predominant in the visible light. If one replaces the single perovskite layer with a perovskite/IV-VI group semiconductor bilayer in the solar cell, the absorption efficiency of the cell should be improved remarkably. After some efforts, we found that GeSe is the best candidate for such a bilayer solar cell. As a consequence, for CH$_3$NH$_3$PbI$_3$/GeSe bilayer structure, the short circuit current density $J_{\text{sc}}$ is dramatically improved, leading to the power conversion efficiency is promoted to $23.77\%$,  being $42.7\%$ larger than that of the conventional perovskite solar cell within the same calculation. This present study may shed new light on developing perovskite solar cells with high efficiency and good performance.

\section{Acknowledgment}

We gratefully thank Qing-Rong Zheng, Qing-Bo Yan and Zheng-Chuan Wang for instructive discussions. This work is supported by Hundred Talents Program of The Chinese Academy of Sciences and the NSFC (Grant No. 11674317). G. S. is supported in part by the MOST (Grant No. 2013CB933401), the NSFC (Grant No. 11474279) and the CAS (Grant No.XDB07010100).

% \bibliography{refs-transport}


%-----------------------------------------------------------------------
%   END DOCUMENT
%-----------------------------------------------------------------------
\end{document}